\documentstyle[aps,epsfig]{revtex}
\begin{document}
%\draft

\title{ Role of defects in superconducting wires for degradation and training }

\author{Narayan Bhattacharya\thanks{E-mail:narayan@veccal.ernet.in}}
\address{Variable  Energy  Cyclotron  Centre,  1/AF Bidhan Nagar,
Kolkata 700 064, India}
\date{\today }
\maketitle
\begin{abstract}

      Intrinsic and extrinsic defects play an important role in training and degradation of superconducting wires and magnets.  Various defects along with experimental data regarding the hierarchy of defects have been presented. Duxbury-Leath model of inverse worst defect cluster has been hypothesized to explain the statistical nature of the phenomenon. `Loss of memory' in the case of training has been explained for the first time. 



\end{abstract}


%\eject

\section{Introduction}
\label{sectoin1}

      In an earler work \cite{r1} it has been shown that `quench' in a superconducting wire can be regarded as a probabilistic phenomenon due to breakdown at local defects of the wire \cite{r2}. Degradation in superconducting magnet or wire can be understood as a `size effect' on the basis of failure probability in percolation model of breakdown and finally training phenomenon was explained for the first time on the basis of breakdown at local defects which are distributed statistically and are based hierarchially (inverse worst defect cluster model) following the lines of Duxbury and Leath statistics. But one of the most important features of `training' is the `loss of memory' behaviour which were not explained in the earlier model. If a superconducting magnet is trained, it remains so, as long as it is immersed in liquid helium at $4.2^{0}K$ and a transport current is passing through it. As soon as the magnet is taken out of liquid helium bath and kept at room temperature for quite sometime and then it is again immersed in liquid helium for energising the magnet, it shows the training behaviour as in earlier case. Hence `training' does not produce any permanent change in the magnet but the `training' is repeated again and again. The motivation of the present work is to explain this particular behaviour of training which was not discussed in the earlier work. Secondly the hierarchial nature of the defects was just proposed in the model intuitively but no experimental proof was cited for the purpose. In this paper an attempt has been made to look into the hierarchial nature of defects from an experimental point of view.

\section{Experimental observation for hierarchy of defects }
\label{section2}

      The implicit assumption of flux pinning studies is that the pinning force or critical current density is determined by the microstructure. Recent work indicates that many practical conductors have transport current densities that do not reach the `intrinsic limits' of the microstructure-flux lattice interaction. Instead, grosser, more microscopic irregularities of the superconducting filaments interrupt the transport current flow, producing `extrinsic limitation' of the critical current. Hence in order to improve the undertstanding of flux pinning microstructure relationship (where $\alpha-Ti$ precipitate play a very important role in the case of $NbTi$ composites) it is clearly important to know whether the samples being studied have critical currents $I_c$ determined by intrinsic flux pinning process \cite{r3} or whether $I_c$ is determined by extrinsic sausaging effects. Many features can introduce sausaging or necking in superconducting composites, reaction between bronze and $Nb$ in $Nb_3Sn$ composite or between $Cu$ and $Nb-Ti$ in $NbTi$ composites can produce hard nodules which are embedded in the superconductor filaments, leading to filament necking during subsequent wire drawing. Chemical or grain structure inhomogenities may cause irregular deformation of the filaments. A particularly attractive way of determining the presence or absence of significant sausaging is by an analysis of resistive transition curve shape \cite{r4}. On the early stages of the transition, the curve can be expressed by a power law given by $V \propto I^n$ \cite{r4}. The index n is a measure of the sharpness of the resistive transition. The sharper the transition, the larger is the n. A very useful diagnostic tool for detecting the presence of the extrinsic limitation is obtained by plotting the n-value versus the magnetic field H which is shown in fig.1 \cite{r4}. When the n-H plot has a steep, approximately uniform negative slope, the composite is operating close to its intrinsic limit as shown in plot (a) of fig.1. As filament non uniformities become important in limiting $I_c$, the n value drops at lower magnetic fields, yielding a plateau shape for the n-H plot (b). In very badly sausaged composites, n-H plateau occurs at very low  and extends over a large range of magnetic fields (plot (c)). In such a case the n-value (5 Tesla) might be 10-30 and the composite $I_c$ can be said to be extrinsically limited. In the intrinsic limit n can exceed 80 at 5 Tesla. It is shown that the change from intrinsic to extrinsic behaviour which is dominated by hierarchy of defects is quite gradual for $NbTi$ composites. This proves that there is no single well defined value of critical current. In fact the transition from the flux pinning (zero resistence state) to the flux flow (resistive) state extends over a range of current values which is nothing but the manifestation of training. 

     Larbalestier et.al. \cite{r5} has shown experimentally, how different degrees of heat treatment at same temperature produce different degrees of sausaging by the growth of intermetallic particles like $CuTi$ and how they degrade the conductor (table1).        
        
\begin{table}
\caption{Effect of heat treatment time and number on the $J_c$ of laboratory produced composite at $375^{0}C$.}
\bigskip
\begin{tabular}{ccccc}
Composite & Filament dia. & Heat treatment & $J_c$ (5T) & $J_c$ (8T)  \\
&&schedule &&\\
 & $\mu m$ & & $A/mm^2$ & $A/mm^2$ \\ \hline
          &       &                       &         &         \\ 
   CB 1210&  21 & 6$\times$10hrs & 2705 & 1145\\
              &       &                       &         &         \\
   CB 1710&  4.0&                      & 2640 & 1080\\
              &  3.2&                      & 2265 &   995\\
           &       &                       &         &         \\
  CB 1230&  19 & 6$\times$40hrs & 3190 & 1135\\
           &       &                       &         &         \\
  CB 1730&  5.3&                      & 1920 &         \\
              &  3.2&                      & 1320 &  510\\
           &       &                       &         &         \\
  CB 1233&  5.3& 9$\times$40hrs & 1590 & 1020\\
              &  4.5&                      &   935 &   324\\

\end{tabular} 
\end{table}
\nopagebreak

      Ekin et.al. \cite{r6} have measured the average filament diameter $<D>$ and averaged diameter fluctuation $<\Delta D>$ for samples A,B and C and have shown clearly how the n value increses progressively with the decrease in the $<\Delta D>$ (table2). Their experiments also show that there is a heirarchy of sausaging according to  $<D>$. Hence within the same filament or composite, in accordance with various degrees of sausaging at various locations the critical current will increase progressively. Table-2 shows the correlations of filament irregularity with n.    

\begin{table}
\caption{Filament diameter variations for three Nb-Ti conductors.}
\bigskip
\begin{tabular}{cccccccccc}
Conductor& Size & Number of & Twist & Cu:Nb-Ti & Average & $D_{max}/D_{min}$& $\Delta D$ &Deformation&n \\
              &        &  filaments  & pitch &              &filament dia. &                      &                  &length&\\
              & (mm) &                & (cm)  &              & $<D>(\mu m)$  &                  & $(\mu m)$   &$(<D>)$&  \\ \hline
    &      &       &        &      &      &       &     &    &  \\
  Sample A&0.51$\times$1.02&187&1.8&4.77&27& 1.33& 8.9& 4&18\\
   &      &       &        &      &      &       &     &    &  \\
  Sample B&0.56$\times$0.71&180&1.3&1.8 &33&  1.20& 6.5& 8&35\\
   &      &       &        &      &      &       &     &    &  \\
   Sample C&1.00 diameter      &114&2.5&5.5 &37& 1.06& 2.2& 6&53\\
 
\end{tabular} 
\end{table}
\nopagebreak

\section{The intrinsic limit}
\label{section3}

      It has been discussed in the beginning that in the absence of extrinsic limit, the critical current density is decided by the intrinsic limit which is dictated by the $\alpha-Ti$ precipitates. These precipitates act as the pinning centres for the $NbTi$ composites. Calculations of the elementary pinning $f_p$ produced by the $\alpha-Ti$ precipitates in $NbTi$ composites indicate that the maximum pinning critical current density $J_c$ should be about $18000 A/mm^2$ at the benchmark field and temperature 5 Tesla and $4.2^{0}K$. This value is almost five times higher than the highest values yet measured which are nearer to $3700 A/mm^2$. This relatively poor experimental performance could have several causes. For example, flux pinning calculations are seldom accurate to the factors of 2 or 3. Also, the size, spacing and volume fraction of $\alpha-Ti$ may not be optimum and in round wires there is no preferential alignment of $\alpha-Ti$ ribbons with the vortices. Bulk pinning force $F_p$ can indeed be made stronger and $J_c$ increased by refining the $\alpha-Ti$ precipitate structure by cold working. $F_p$ is given by      

\begin{equation}
 F_p = N_p.\lambda.f_p,
\label{seqn1}
\end{equation}
\noindent
where $\lambda$ is the efficiency factor, $f_p$ is the elementary pinning force and $N_p$ is the precipitate density. $J_c$ has been found to increase in direct proportion to the volume fraction of precipitate \cite{r6}. Hence the only parameter which can change for a particular composite is the efficiency parameter $\lambda$ which is decided by the alignment of the microstructure. Cooley et. al. has shown that \cite{r7} by cold rolling the composite the aspect ratio can be changed and $J_c$ values can be increased with the increase in the aspect ratio (table3). 

\begin{table}
\caption{Mean filament thickness, cross-sectional area, aspect ratio and transport critical current density (at 5 T, $4.2^0K$) for the rolled monofilament samples. The aspect ratio of 1.0 corresponds to the round wire prior to rolling ($\epsilon_{f} = 4.17$).}
\bigskip
\begin{tabular}{cccc}
Thickness & Area & Aspect ratio & $J_c$  \\

$mm$ & $mm^2$ &     & $A/mm^2$ \\ \hline
          &       &        &         \\ 
 0.153 & 0.0188 & 1.00 & 2610 \\
              &       &        &         \\
 0.129 &  0.0192 & 1.16 & 2760 \\
           &        &          &         \\
 0.102 &  0.0188 & 1.81 & 3090 \\
           &       &         &         \\
 0.083 &  0.0186& 2.67 & 3230 \\
           &       &        &         \\
 0.062 &  0.0175& 4.57  & 3880 \\
           &       &       &         \\
 0.046 &  0.0164& 7.82 &4880 \\
           &       &       &         \\
 0.040 &  0.0152& 9.73 &5200 \\
           &       &       &         \\
 0.037 &  0.0150& 10.9 &4390 \\
           &       &         &         \\
 0.035 &  0.0147& 12.1 &4080 \\
           &       &      &         \\

\end{tabular} 
\end{table}
\nopagebreak

\section{Loss of memory}
\label{section4}

      Now it is well established experimentally how by changing the alignment of $\alpha-Ti$ precipitates the efficiency factor $\lambda$, for the bulk pinning force can be changed. Hence even without changing $f_p$ and $N_p$, the critical current density can increase according to the degree of alignment of $\alpha-Ti$ precipitate in the superconducting composite. According to our previous model, one can argue that the hierarchy of defects is decided by the degree of alignment of $\alpha-Ti$ precipitate. As more and more $\alpha-Ti$ precipitates are aligned, the critical current progressively increases and reaches plateau when $\alpha-Ti$ precipitates are optimally aligned. But when the magnet is not in use i.e. taken out of liquid helium cryostat and there is no transport current passing through it, because of thermal relaxation, all the $\alpha-Ti$ precipitates are relaxed and randomly oriented. Hence it can be argued that the training of the superconducting magnet is a temporary phenomenon which can be easily destroyed by thermal excitation. Each time the magnet is energised, it shows training which is an intrinsic behaviour of the superconducting magnet. Now it is clear that even with all possible precautions of wire-motion and epoxy-cracking, the basic behaviour of training in superconducting magnets can not be avoided altogether. Warnes \cite{r8} has proposed a model in which the critical current distribution along the length of a superconducting composite has been correlated to the shape of the $\partial^2V/\partial{I^2}$ for critical current transition shape. According to this model there is a variation in critical current capacity of a filament along its length. When the current exceeds the critical current of a particular section of the filament, voltage is developed across the composite, as a result of the extra current being transferred to the resistive matrix. The model predicts that the second derivative $\partial^2V/\partial{I^2}$ of the $V-I$ transition yields the distribution of the critical current in the wire sample. The distribution can be due to intrinsic or extrinsic limitation along the length of the filament. If there is a sausaging or necking in the sample, then there will be a filament cross-section distribution along the length of the wire. 

\section{Experimental verification of the inverse worst defect cluster model or the multivalued weak link (MVWL) model}
\label{section5}

      In this experiment experiment, $V(I)$ was measured in the undeformed condition and was then mechanically deformed. The pinning centres come closer and become more aligned to the applied magnetic field due to deformation and more effective in holding the flux line. In the $\partial^2V/\partial{I^2}$ curve we see an extra peak indicating critical current at a higher value of current compared to the undeformed measurements. This indicates that the critical current at the deformed places actually increases with respect to the original undeformed wire \cite{r8}. 
  
\section{Summary and conclusion}
\label{section6}

      It is concluded that movements of the of the dislocation defects in the stress field causes some work done  and as a result of which some heat is generated. Hence even if all the precautions are taken for eliminating wire motion  and epoxy cracking, beacause of migration of defects there will be work done and heat will be generated which may ultimately cause quenching depending on the cooling conditions.

\begin{figure}[h]
%\eject\centerline{\jpgfig{file=ncb-image.jpg,height=14cm,width=14cm}}
%\eject\centerline{\epsfig{file=fig1.eps,height=14cm,width=14cm}}
\caption
{The n-H plot for a series of wires with increasing extrinsic limitations to the $I_c$. The composites shown are (a) intrinsically limited; (b) mildly sausaged; (c) badly sausaged. (Please see 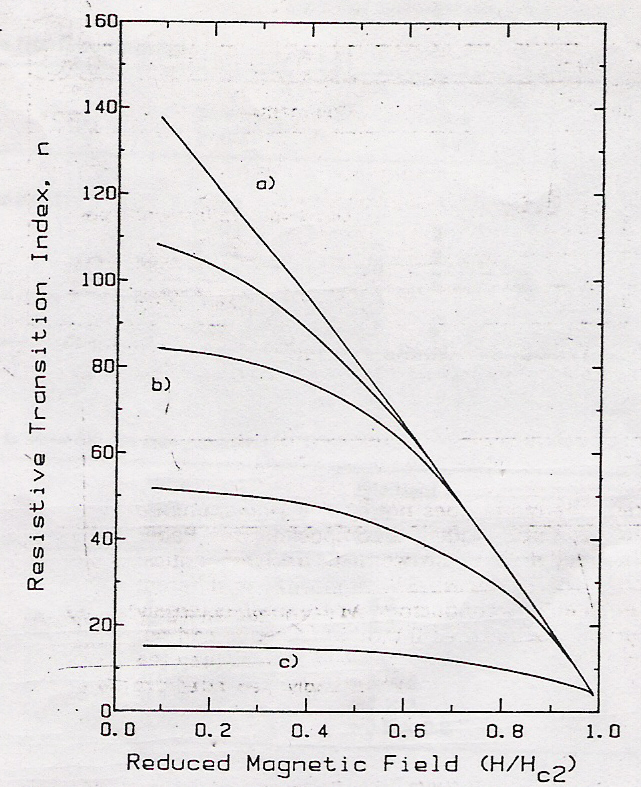) }
\label{fig1}
\end{figure}

\end{document}